\documentclass[twocolumn,prl,showpacs,superscriptaddress]{revtex4}
\usepackage{graphicx}

\usepackage[colorlinks,bookmarks=false,citecolor=blue,linkcolor=red,urlcolor=blue]{hyperref}

\begin{document}
\title{Field-Induced Gap in a Quantum  Spin-1/2 Chain in a Strong Magnetic Field}

\author{S.A. Zvyagin}
\affiliation{Dresden High Magnetic Field Laboratory (HLD),
Forschungszentrum Dresden-Rossendorf (FZD), 01314 Dresden, Germany}
\author{E.~\v{C}i\v{z}m\'{a}r}
\affiliation{Centre of Low Temperature Physics, P.J. \v{S}af\'{a}rik
  University, SK-041 54 Ko\v{s}ice, Slovakia}
\author{M.~Ozerov}
\affiliation{Dresden High Magnetic Field Laboratory (HLD),
Forschungszentrum Dresden-Rossendorf (FZD), 01314 Dresden, Germany}
\author{J.~Wosnitza}
\affiliation{Dresden High Magnetic Field Laboratory (HLD),
Forschungszentrum Dresden-Rossendorf (FZD), 01314 Dresden, Germany}
\author{R. Feyerherm}
\affiliation{Helmholtz-Zentrum Berlin f\"ur Materialien und Energie, 14109 Berlin, Germany}
\author{S.R. Manmana}
\affiliation{JILA, Department of Physics, University of Colorado, 440 UCB, Boulder, Colorado 80309, USA}
\affiliation{Institute of Theoretical Physics, Ecole Polytechnique F\'ed\'erale de Lausanne, CH-1015 Lausanne, Switzerland}
\author{F. Mila}
\affiliation{Institute of Theoretical Physics, Ecole Polytechnique F\'ed\'erale de Lausanne, CH-1015 Lausanne, Switzerland}

\begin{abstract}
Magnetic excitations in copper pyrimidine dinitrate, a spin-1/2 antiferromagnetic chain  with alternating $g$-tensor and Dzyaloshinskii-Moriya interactions that exhibits a field-induced spin gap, are probed by means of pulsed-field electron spin resonance  spectroscopy.  In particular, we report on a minimum of the gap in the vicinity of the saturation field $H_{sat}=48.5$ T associated with a transition from the sine-Gordon region (with soliton-breather elementary excitations) to a spin-polarized state (with magnon excitations). This interpretation is fully confirmed by the quantitative agreement over the entire field range of the experimental data with the DMRG investigation of the spin-1/2 Heisenberg chain with a staggered transverse field.
\end{abstract}

\maketitle

\emph{Introduction.---}
Due to recent progress in theory and the growing number of physical realizations, low-dimensional quantum magnets continue to receive a considerable amount of attention. They serve as model systems for investigating numerous fascinating phenomena in materials with cooperative ground states, in particular, induced by high magnetic fields.  The way a magnetic field changes the ground-state properties and, correspondingly,  the low-energy excitation spectrum of low-dimensional magnets is one of the fundamental aspects in quantum magnetism. For example, the zero-field ground state of an isotropic $S=1/2$ Heisenberg antiferromagnetic (AF) chain with uniform nearest-neighbor exchange coupling is a spin singlet, and its spin dynamics is determined by a gapless two-particle continuum of fractional $S=1/2$ excitations, called spinons.
Application of an  external magnetic field $H$ leads to a pronounced rearrangement of the excitation spectrum, making the soft modes incommensurate \cite{Mueller,Stone} but leaving the spinon continuum gapless.
However, in a fully spin-polarized phase, $H > H_{sat}$, the excitation spectrum is gapped and  dominated by ordinary spin waves (magnons). The low-energy excitation spectrum of such an ideal isotropic Heisenberg $S=1/2$ chain  can be described in terms of an  effective free massless boson theory \cite{Luther,Korepin,Lukyanov}.

The presence of additional interactions such as Dzyaloshinskii-Moriya (DM) interaction\cite{DMoriginal} can  significantly alter the physical  properties of such spin systems (see for instance Ref. \cite{Penc}), and,  in particular, their high-field behavior. A Heisenberg spin-1/2 chain with exchange interaction $J$ and DM interaction (or alternating $g$-tensor)  in a field $H$ can be mapped to a simple Heisenberg chain with a staggered transverse field $h \propto H$ \cite{Essler99,OshikawaAffleck97,AffleckOshikawa99prb}, described by the effective spin Hamiltonian
\begin{equation}
\mathcal{H} = \sum_j \left[ J \mathbf{S}_j \cdot \mathbf{S}_{j+1} - H S_j^z - h \, (-1)^j \, S_j^x \right].
\label{eq:microscopicHam}
\end{equation}

The sine-Gordon quantum field theory \cite{OshikawaAffleck97,Essler99,AffleckOshikawa99prb} applied for this model predicts that the elementary excitation spectrum is to be  formed by  solitons and their multiple bound states (breathers) with an energy gap $\Delta \propto H^{2/3}$ formed by  the first breather mode \cite{Zvyagin_solitons,Nojiri}. Such a sine-Gordon spin model has been found to be realized in a number of compounds
\cite{Dender,Asano,Feyerherm,Kohgi,Kenzelmann,Umegaki}. Among others, copper benzoate \cite{Dender} and copper pyrimidine dinitrate \cite{Feyerherm} are the most intensively studied materials.

The predictions of the sine-Gordon theory are limited to the range of small to moderate fields.
Numerical simulations based on the Density Matrix Renormalization Group (DMRG) \cite{DMRGpapers} by Zhao et al. \cite{Zhao} have shown that when the field approaches the
saturated phase the energy gap is a non-monotonous function of the field, and that
it presents a minimum around the saturation field before the linear increase above saturation.
This result has been understood in analytical terms somewhat later by Fouet et al. \cite{Fouet}, who have shown,
using field-theory arguments, that the gap around the saturation field in one-dimensional spin systems scales as $h^{4/5}$. This exponent
is larger than the exponent that controls the gap opening at low field, and for a small enough proportionality
constant between the staggered field and the external field, this leads to a local minimum of the gap
around the saturation field.

These remarkable changes in the excitation spectrum in the vicinity of the saturation field ($H_{sat} \approx 27$ T) have been identified experimentally first in copper benzoate   \cite{Nojiri}.
The data obtained for $H\|c$ was compared to DMRG calculations performed for this material for $H\|c''$ \cite{Zhao}. The results are in qualitative agreement, but the difference between experiment and theory
at high field is significant, both regarding the position of the minimum and the magnitude of the gap at high fields.
Since the field orientations were different in the experiment and the calculation, it is not possible
to decide on the data at hand whether the microscopic model of Eq.~(\ref{eq:microscopicHam}) can accurately describe the high-field behavior of copper benzoate (and other sine-Gordon quantum magnets), a question that has remained unanswered since then.

Recently, the ESR excitation spectrum in the quantum sine-Gordon $S=1/2$ AF chain material copper pyrimidine dinitrate (hereafter Cu-PM) \cite{Feyerherm} had been studied in magnetic fields up to 25 T  \cite{Zvyagin_solitons,Zvyagin_pertr}. A signature of soliton and three breathers, as well as specific temperature and field dependences of ESR parameters (linewidth and $g$-factor) beautifully confirmed the applicability of the  sine-Gordon quantum-field theoretical approach for Cu-PM for $H < J/g\mu_B$.  Here, we extend the range of magnetic fields used in ESR experiments  up to 63 T. We report on pronounced changes in the ESR spectrum in the vicinity of the saturation field, $H_{sat}=48.5$ T \cite{Wolter}, clearly indicating a transition from the soliton-breather to the magnon state.
Comparison of the experimental data with numerically exact DMRG results based on the microscopic model described with Eq.~(\ref{eq:microscopicHam})  revealed excellent {\it quantitative} agreement.

\begin{figure}[h]
\begin{center}
\includegraphics[width=0.45\textwidth]{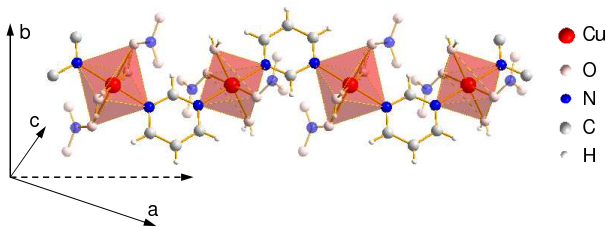}
\caption{\label{fig:structure} (Color online) Schematic view of the crystal structure of Cu-PM, showing the alternation of octahedrons. The Cu-chains are
running parallel to the short $ac$ diagonal (denoted by the dashed arrow).
}
\end{center}
\end{figure}

\emph{Experimental.--- }
Cu-PM, [PM-$\rm Cu(NO_{3})_{2}(H_{2}O)_{2}$]$_{n}$
 (PM = pyrimidine),
crystallizes in a monoclinic structure belonging to the space group
$C2/c$ with four formula units per unit cell \cite{Feyerherm}. The
lattice constants obtained from  single-crystal X-ray
diffraction are $a=12.404$~\AA, $b=11.511$~\AA, $c=7.518$~\AA,
$\beta=115.0^{\circ}$. The Cu ions form  chains
running parallel to the short $ac$ diagonal (Fig.\ \ref{fig:structure}). The Cu ions are
linked by the N-C-N moieties of pyrimidine, which constitute the
intrachain magnetic exchange pathway. The Cu coordination sphere is a distorted
octahedron, built from an almost squarish  N-O-N-O equatorial plane
and two oxygens in the axial positions. In this approximately
tetragonal local symmetry, the local principal axis of each
octahedron is tilted from the $ac$ plane by $\pm 29.4^{\circ}$.
Since this axis almost coincides with the principal axis of the
$g$-tensor,   the $g$-tensors for neighboring Cu ions are
staggered. The exchange constant $J=36\pm0.5$~K was extracted from
the single-crystal susceptibility \cite{Feyerherm} and confirmed
by magnetization measurements \cite{Wolter}.

High-field ESR experiments of Cu-PM were performed  at the Dresden High Magnetic Field Laboratory (Hochfeld-Magnetlabor Dresden, HLD)  using a pulsed-field
ESR spectrometer \cite{FEL_zvyagin} equipped with  VDI  sources of millimeter-wave   radiation (product of Virginia Diodes Inc.) and with a transmission-type probe in the Faraday configuration.    A 8.5 MJ/70 T magnet was employed  to generate  pulsed magnetic fields with a pulse-field rise time of
35 ms and  full-pulse durations of about $150$ ms. The magnetic field was
applied along the $c''$ direction, which is characterized by the
maximal value of the staggered magnetization for Cu-PM
\cite{Feyerherm}. Experiments were performed  at a temperature  of 1.9 K.  High-quality single-crystals of
Cu-PM with typical  size of 3x3x0.5 mm$^3$ were used.  2,2-Diphenyl-1-Picrylhydrazyl (known as DPPH) was employed for the calibration of the magnetic field.

\begin{figure}[h]
\begin{center}
\includegraphics[width=0.45\textwidth]{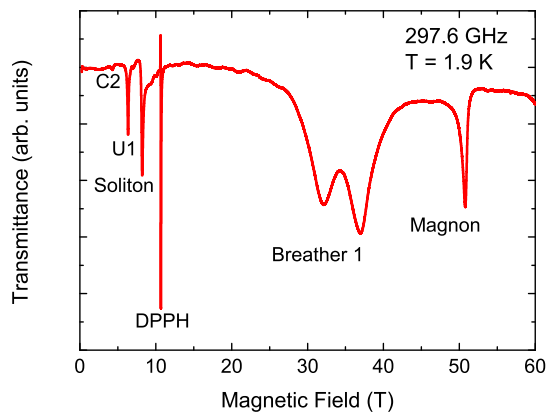}
%\vspace{-0.5cm}
\caption{\label{fig:spectrum} (Color online) ESR transmission spectrum of Cu-PM, taken at a frequency of 297.6 GHz at $T=1.9$ K (DPPH is used as a marker).
}
\end{center}
\end{figure}

\begin{figure*}[th]
\begin{center}
\includegraphics[width=0.8\textwidth]{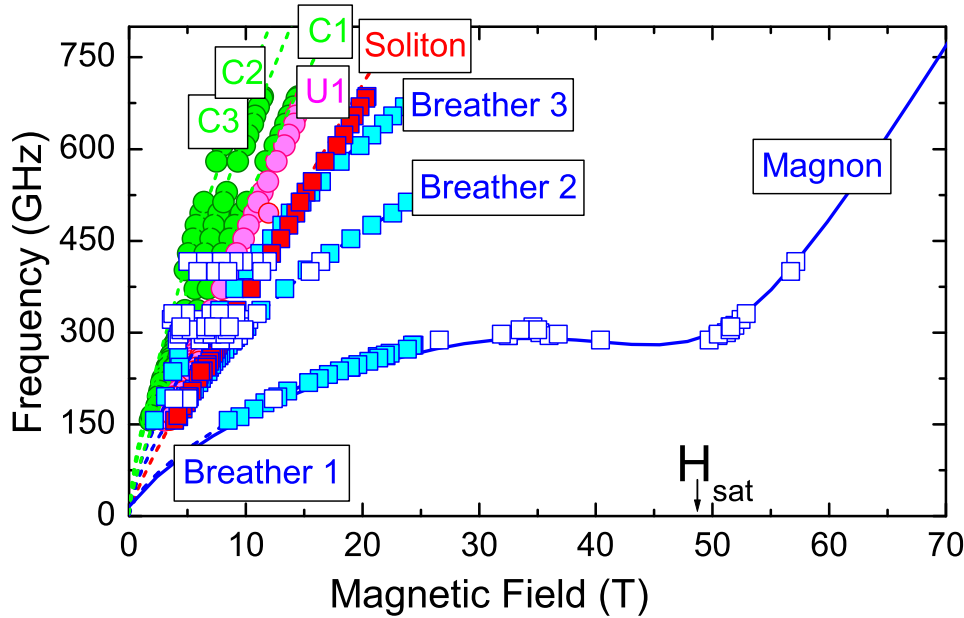}
%\includegraphics[width=0.9\textwidth]{Fig3_FFD}
%\vspace{-0.5cm}
\caption{\label{fig:FFD} (Color online) Frequency-field dependence of the detected ESR modes in Cu-PM.  Experimental data are denoted by symbols, and lines correspond to the results by use of the sine-Gordon theory (dashed lines) and of the DMRG (solid line). Data denoted by closed symbols  are taken from Ref.~[\onlinecite{Zvyagin_solitons}], while  open symbols are experimental results obtained in the present study.
}
\end{center}
\end{figure*}

\emph{Results and discussion.---}
In Fig.\ \ref{fig:spectrum}, we show a typical  pulsed-field  ESR spectrum taken in Cu-PM at  1.9 K and at a frequency of  297.6 GHz. Several very pronounced ESR modes were resolved. As mentioned above, ESR spectra in Cu-PM were previously measured in magnetic fields up to 25 T \cite{Zvyagin_solitons}. The corresponding data from Ref. \cite{Zvyagin_solitons} and results of the present pulsed-field ESR experiments are compiled in Fig.\ \ref{fig:FFD} using  filled and open symbols, respectively. In addition to the soliton and three breather modes, additional ESR modes were observed in the magnetic  excitation spectra, including the ones labeled by $C1-C3$ (which correspond to the edges of the soliton-breather continua) and the mode $U1$, which can be related to bound states due to topological edge effects \cite{Lou1,Lou2}.

According to the  sine-Gordon quantum field
theory for  quantum spin-1/2 chains \cite{OshikawaAffleck97,AffleckOshikawa99prb}, the field-induced energy gap can be calculated using the expression for  the first breather excitation mode  \cite{AffleckOshikawa99prb}:
\begin{equation}
\label{bregap} \Delta_{g}=2J {2\Gamma({\xi\over2})v_{F} \over
  \sqrt{\pi}\Gamma(\frac{1+\xi}{2})}
\left[ {g\mu_{B}H \over J v_{F}} {\pi\Gamma({1\over 1+\xi}) c A_{x} \over
  2\Gamma({\xi\over 1+\xi})}\right]^{\frac{1+\xi}{2}}\sin(\pi\xi/2). \quad
\end{equation}
In this expression, $c$ is the proportionality coefficient connecting the uniform
applied field $H$ and the effective staggered field $h=cH$ ($c=0.083$ for Cu-PM \cite{Zvyagin_pertr}), the
parameter $\xi=\big(2/(\pi R^{2})-1\big)^{-1}$, where $R$ is the
compactification radius, and $v_{F}$ has the meaning of the Fermi
velocity \cite{AffleckOshikawa99prb}.
The amplitude $A_{x}$, which is also a function of $H$, has been recently computed
numerically \cite{EsslerFH}. Excellent agreement between  experimental data and results of the calculation using  Eq. (\ref{bregap}) for the first breather mode in magnetic fields up to 25 T was found  \cite{Zvyagin_solitons}.

Since the  expression~(\ref{bregap}) is valid for a wide range of fields but only up to $H\sim J/g\mu_{B}$,
it is necessary to turn to another approach to calculate the spin gap in the microscopic model of Eq.~(\ref{eq:microscopicHam}) at high fields ($H > J/g\mu_{B}$). To this end, we have performed DMRG calculations
directly for the model of Eq.~(\ref{eq:microscopicHam}) since this numerical technique is very accurate
for all fields.  The results of the DMRG  calculations for the field-induced gap using the microscopic model described by Eq.~(\ref{eq:microscopicHam}) with $L=200$ lattice sites and $c=0.083$ are shown in Fig.\ \ref{fig:FFD} by a solid line.
These results were obtained by performing up to 10 DMRG sweeps and keeping up to $m=500$ states, resulting in a discarded weight $\varepsilon \ll 10^{-10}$. Note that due to the presence of the transverse-field term in the Hamiltonian~(\ref{eq:microscopicHam}), $S^z_{\rm total}$ is not a good quantum number, limiting the possible system sizes. However, already for the system size shown, the results for the gap are essentially the same as in the thermodynamic limit, as revealed by comparing with smaller systems.
Several remarks are in order. First of all, the DMRG results are in very good agreement with the prediction
of the sine-Gordon theory in its limit of validity (low field). Secondly, and more importantly, the agreement
between the DMRG results and the experimental results is excellent at all values of the field, in particular up to $H_{sat}$ and above. This establishes that the microscopic model of Eq.~(\ref{eq:microscopicHam}) provides a quantitative
description of the spin-gap behaviour in spin-1/2 sine-Gordon chains for all values of the field.

Now  that the validity of the theoretical description has been established, let us comment on the physical
mechanism behind the non-monotonic increase of the gap \cite{Zhao,Fouet}. In the absence of a staggered field, the system is
gapless below $H=H_{sat}$, and it is in the fully spin-polarized state above. In that state, elementary
excitations are magnons, and the gap opens linearly with $H-H_{sat}$. The presence of a staggered field
perpendicular to the external field opens a gap in the spectrum because it breaks the rotational symmetry
around the field. Now, the impact of the staggered field is related to the magnitude of the transverse
magnetization it induces. Close to saturation, the spins are almost polarized, and the system cannot
develop a large transverse staggered magnetization. So the staggered field is less efficient to open
a gap close to saturation than at low field. This is the basic mechanism behind the different scalings
of the gap with $c$ at low field ($\Delta \propto c^{2/3}$) and close to saturation ($\Delta \propto c^{4/5}$).
For small enough $c$, this leads to a minimum of the gap around the saturation field. This explains the small
but still well-resolved dip in the frequency-field dependence of magnetic excitations in Cu-PM in the vicinity of $H_{sat}$. Such a behavior appears to be a general  feature of the high-field excitation spectrum of  quantum spin-1/2 chain systems with  alternating $g$-tensor and/or Dzyaloshonskii-Moriya interactions.

\emph{Summary.--- }
We have presented a detailed ESR studies of Cu-PM, a material containing $S=1/2$ AF
chains with alternating $g$-tensor and DM interaction, and exhibiting a
field-induced gap.  From that we extracted the field-dependent ESR excitation spectrum of Cu-PM in magnetic fields up to 63 T. The field-induced change in the gap behavior  was observed
directly, clearly indicating the effect of a suppression of the soliton-breather magnetic-excitation regime by  a strong magnetic field followed by a transition into the fully spin-polarized phase with magnons as elementary excitations.  By comparing the entire set of data
with results of  DMRG calculations (based on parameters obtained earlier \cite{Zvyagin_solitons,Zvyagin_pertr}) the validity of the used theoretical approach has been proven.   Excellent agreement between the experiment and results of the calculations was found. Our results are relevant for the understanding of the
spin dynamics in copper benzoate and other $S=1/2$
Heisenberg antiferromagnetic chain systems.

\emph{Acknowledgments.--} The authors would like to thank  A.K. Kolezhuk for
fruitful discussions. This work was partly supported by the Deutsche
Forschungsgemeinschaft and EuroMagNET (EU contract No. 228043).
E.\v{C} appreciates the financial support by VEGA 1/0078/09 and APVV-0006-07, and S.R.M. acknowledges the support by PIF-NSF (grant No. 0904017). F.M. acknowledges the  support of the Swiss National Fund
and of MaNEP. We acknowledge the allocation of CPU time on the pleiades-cluster at EPFL.

\end{document}